\documentstyle[11pt,paspconf]{article}

\begin{document}
\centerline{\null}
\vskip-1.0truein
\noindent
To appear in {\it Galactic and Cluster Cooling Flows},
ed.\ by N. Soker,
(San Francisco: Publ.\ Astr.\ Soc.\ Pacific), in press.
\vskip0.7truein

\title{Cluster Cooling Flows: Recent Progress and
	Outstanding Questions}

\author{Craig L. Sarazin}
\affil{Department of Astronomy, University of Virginia,
	Charlottesville, VA 22903-0818 U.S.A.}

\begin{abstract}
Several recent results on cluster cooling flows are discussed, and
a number of remaining mysteries are described.
Observations of excess soft X-ray provides the only direct evidence
for a major repository for the cooled gas.
Unfortunately, the frequency of occurrence of large excess columns is
uncertain.
Excess absorption would affect the interpretation of most X-ray
observations of cooling flows.
There is also great uncertainty about the physical state of the
material producing the absorption.
Radio observations have ruled out most forms of cold gas.
Unfortunately, the theoretical models which have been constructed
for the absorbing clouds give very discrepant results, and we have
no generally accepted model for the possible physical state of the
clouds.

Recent radio observations show that the plasma in cooling flows is
strong magnetized.
Models for cooling flows including the dynamical effects of the magnetic
field are needed.
We also need to understand what ultimately happens to the advected magnetic
flux.
At present, it appears most likely that the field is by field
line reconnection.

The radio sources associated with the central galaxies in cluster
cooling flows seem to come in two varieties.
There are ``lobe-dominated sources'' with strong radio jets and
well-defined radio lobes, and ``amorphous'' sources without strong
jets or lobes.
In the lobe-dominated sources, the radio pressures agree approximately
with the pressures of the X-ray emitting thermal gas, the X-ray
and radio images anticorrelate, and the radio sources are highly
polarized and have very large Faraday rotations.
In the amorphous sources, the radio pressures are much smaller than
the X-ray pressures, the X-ray and radio images correlate, and the radio
sources are strongly depolarized.
These results suggest that the lobes in the lobe-dominated sources
have displaced and are confined by the X-ray emitting gas, while
in the amorphous sources, the radio and thermal plasma are mixed.
\end{abstract}


\section{Introduction} \label{sec:intro}

Observations with the $Einstein$ Observatory and some earlier observations
established that large quantities of gas are cooling below
X--ray emitting temperatures in the cores of many clusters
(see Fabian et al.\ [1984, 1991], and
Fabian [1994] for reviews).
Typical cooling rates are $\sim$100 $M_\odot$ yr$^{-1}$.
I will discuss some recent observations of cluster cooling flows,
and also present some things which I find puzzling.
Of course, the greatest mystery about cooling flows is the existence
and nature of the ultimate repository of the gas seen to cool through
the X-ray band.
If cooling flows are long--lived phenomena,
roughly
$M_{cool} \sim 10^{12} \, M_\odot$
of material would cool over the lifetime of the cluster.

\section{Excess X-ray Absorption} \label{sec:absorp}

\subsection{X-Ray Observations} \label{sec:absorp_obs}

One of the most exciting and potentially important recent discoveries
concerning cooling flows was the detection
of excess X-ray absorption in cluster cooling flows.
Through a re-analysis of $Einstein$ Solid State Spectrometer (SSS)
X-ray spectra of the central regions of cooling flow clusters,
White et al.\ (1991) found evidence for very large amounts of
excess soft X-ray absorption.
The excess column densities of X-ray absorbing material
were typically $\Delta N_H \approx 10^{21}$ cm$^{-2}$.
Excess absorptions have also been found using other detectors on
$Einstein$, $Ginga$, BBXRT, and ASCA
(Johnstone et al.\ 1992;
Lea et al.\ 1982;
Miyaji et al.\ 1993;
Fabian et al.\ 1994).
In several cases, ROSAT PSPC spectral images have shown that the
excess absorption is concentrated to the cooling flow region with
a radius of $\sim$200 kpc.
Thus, the total required mass of cold gas determined by multiplying the
excess column density by the area is about
\begin{equation} \label{eq:mcold}
M_{cold} \approx 1.4 \times 10^{12} \, M_\odot
\left( \frac{ \Delta N_H}{10^{21} \, {\rm cm}^{-2}} \right)
\left( \frac{ r_c}{200 \, {\rm kpc}} \right) \, .
\end{equation}
This is comparable to the total mass expected to cool out of the X-ray band
over a Hubble time.

\begin{figure}[t]
\vspace{2.1in}
\includegraphics{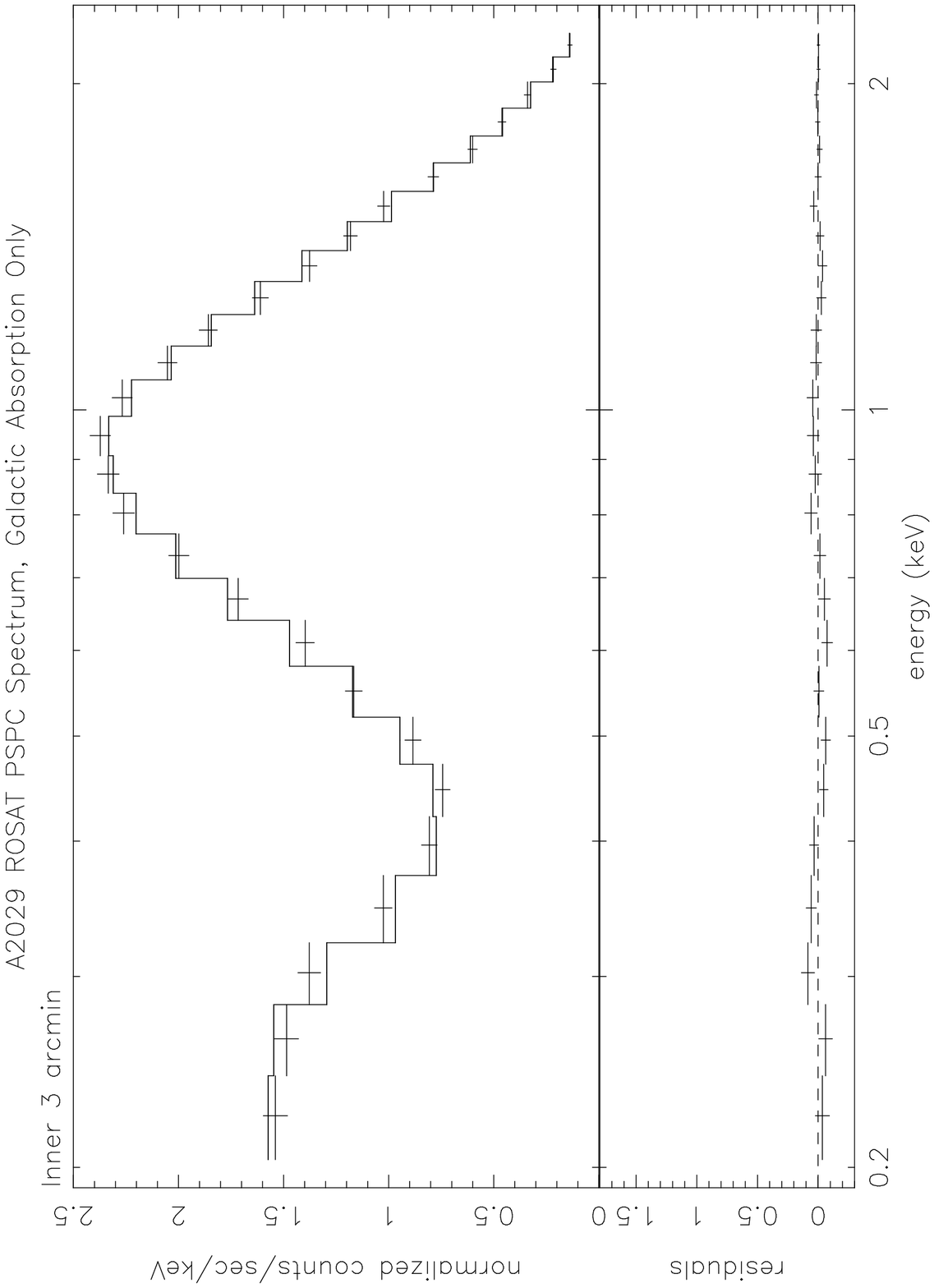}
\includegraphics{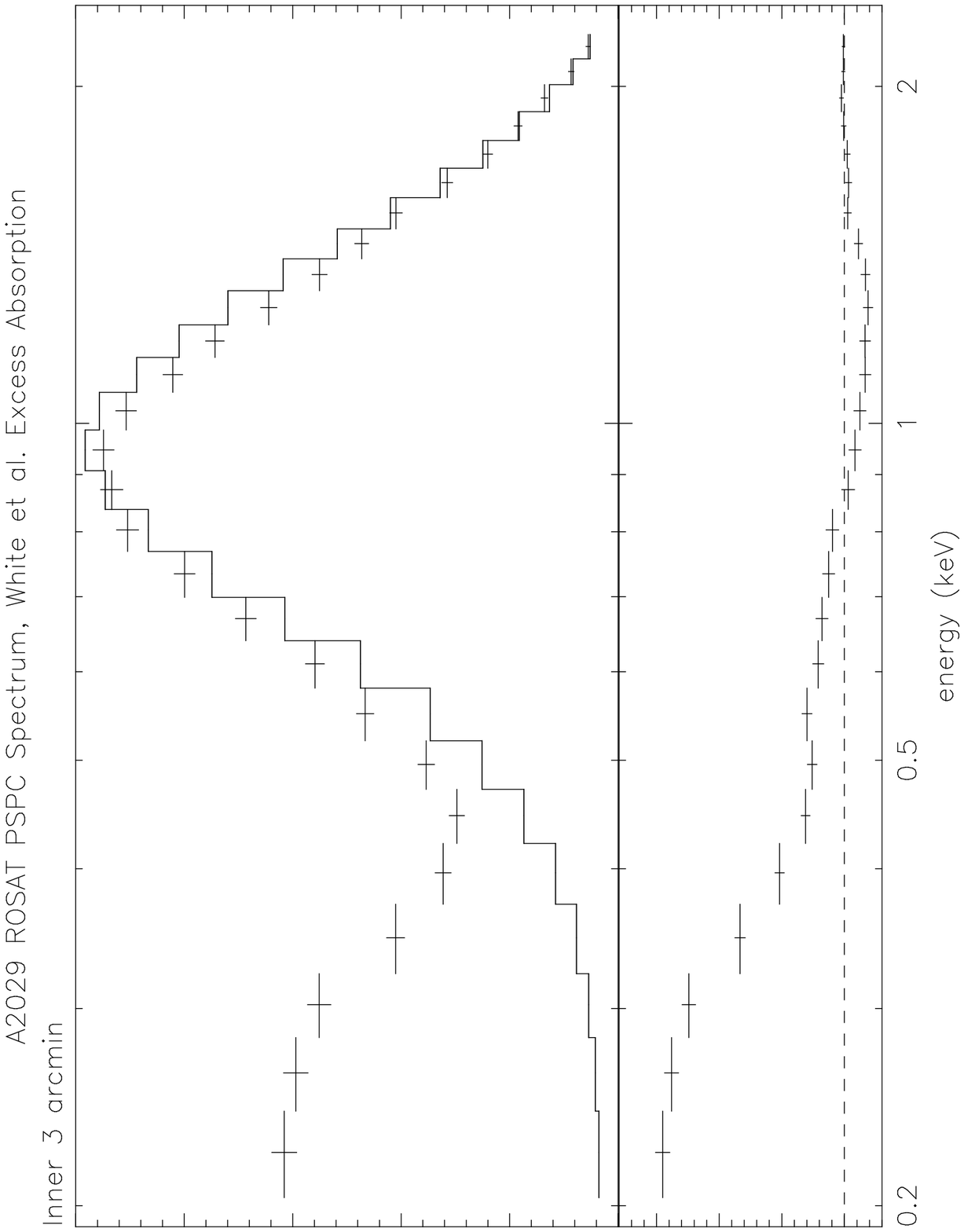}
\caption{The ROSAT PSPC X-ray spectrum for the central 3 arcmin radius
region of the A2029 cluster
(Sarazin et al.\ 1996).
In the left hand panel, the solid histogram is the best-fit single temperature
plus cooling flow model assuming only Galactic absorption.
In the right hand panel, the solid histogram is the best-fit model with
excess absorption fixed at the value from White et al.\ (1991).}
\label{fig:a2029pspc}
\end{figure}

Because of its soft X-ray band, moderate spectral resolution, reasonably
accurate calibration, and bimodal spectral
response,
the ROSAT PSPC is an excellent instrument for detecting excess
soft X-ray absorption.
Based on the common detection of very large excesses absorptions
($\Delta N_H \approx 10^{21}$ cm$^{-2}$)
directly in the observed spectra with the $Einstein$ SSS spectra
by White et al.\ (1991), one would have expected to have found
many such cases with the ROSAT PSPC.
A number of large excess absorptions have been published based
on PSPC data (Allen et al.\ 1993, 1995; Irwin \& Sarazin 1995).
However, many PSPC spectra do not show such large excess absorptions
covering all of the emission toward the central cooling region of
the cluster
(Breen 1996).
The ROSAT PSPC spectrum of Abell 2029 illustrates this difference.
White et al.\ (1991) found an excess column of
$\Delta N_H = 1.8 \pm 0.5 \times 10^{21}$ cm$^{-2}$ covering all
of the emission in the central 3 arcmin radius of this cluster.
Figure~\ref{fig:a2029pspc} shows the ROSAT PSPC spectrum of the
same inner 3 arcmin circle
(Sarazin et al.\ 1996).
In the left panel, the solid line gives the best-fit spectral
model, including a cooling flow.
The right panel shows the best-fit model if an excess absorption
equal to the White et al.\ value is assumed.
No excess absorption is required to fit the spectrum of the total
emission in this region.
The 90\% confidence upper limit from the ROSAT PSPC spectrum
of the total emission with the inner 3 arcmin radius is
$\Delta N_H < 1.2 \times 10^{20}$ cm$^{-2}$, which is more
than an order of magnitude below the value found by
White et al.\ (1991).
In general, large excess columns are found much less commonly in
the ROSAT PSPC spectra of the total emission toward cooling flows
than in the $Einstein$ SSS spectra of the same regions
(Breen 1996).
However, significant excess columns have been found towards the cooling flow
components of the ROSAT PSPC X-ray spectra of many cooling flow
clusters by fitting these components separately
(Allen 1996).

Another possible source of concern is that some of the largest excess
columns have been found toward clusters which are themselves at
large Galactic columns
(Allen et al.\ 1993; Irwin \& Sarazin 1995; Breen 1996).
Obviously, this cannot be a selection effect, since excess absorption
should be easier to detect if the Galactic column is low.
Moreover, some clusters show excess absorption which is not
centrally condensed and/or which may be associated with Galactic
interstellar features
(David et al.\ 1996).

Now that all of the ROSAT PSPC data is public, it would be very useful
to analyze a large and complete sample of cluster cooling flows to
determine their excess absorption.
It would be very useful to know how common large excess absorptions
are, and what the distribution of columns is (e.g., the fraction
of cooling flow clusters with excess columns greater than $\Delta N_H$).
It would also be very good to look for correlations between
the excess absorption $\Delta N_H$ and the cooling rate
$\dot M$, and between the excess absorption and the Galactic
column.

The possible tendency for large excess columns to correlate
with large Galactic columns might be explained if the excess columns were
due in part to Galactic material.
In order to convincingly eliminate this possibility, it would be very
useful to map the Galactic interstellar atomic and molecular material
toward one or two very good cases of cooling flows with excess absorption.
Ideally, these cases would be chosen to have large and unambiguously
determined excess absorption columns and small Galactic columns in
the existing surveys.
The need for detailed mapping of the Galactic ISM in these directions
comes about because the existing surveys (e.g., Stark et al.\ 1992)
have been made with large beams which are widely spaced.
Given the angular size of nearby cooling flows, the best technique
would probably be to map the Galactic H~I with the VLA with a
fairly compact array, and to use a large single disk telescope to
map the galactic CO distribution in the same direction.

The most direct method to establish that the excess absorption is
associated with the cluster and not with our Galaxy is to measure the
redshift of the oxygen K absorption edge.
In principle, this should be possible with ASCA for the brightest cooling
flows at redshifts $z \ga 0.07$, but low energy calibration problems
have made this extremely difficult
(Sarazin et al.\ 1996).

\subsection{Cooling Flow Models with Intrinsic Absorption}
\label{sec:absorp_model}

The concentration of the excess absorption toward the center of
several cooling flow clusters
(Allen et al.\ 1993; Irwin \& Sarazin 1995) and the correspondence between
$M_{cool}$ and $M_{cold}$ suggest that the excess absorber is located
within the cooling flow.
For simplicity, in all existing analyses of X-ray spectra, the absorber
has been treated as a foreground screen in front of the cooling flow.
Of course, a foreground absorber and absorber mixed with the emitting gas
give different spectra and other properties.
Wise \& Sarazin (1996) have calculated models for the X-ray emission
of cooling flows with internal absorption.
We assume that the cold absorbing gas has the same distribution as that
of the gas cooling out of the X-ray temperature band.

\begin{figure}[t]
\vspace{1.92in}
\includegraphics{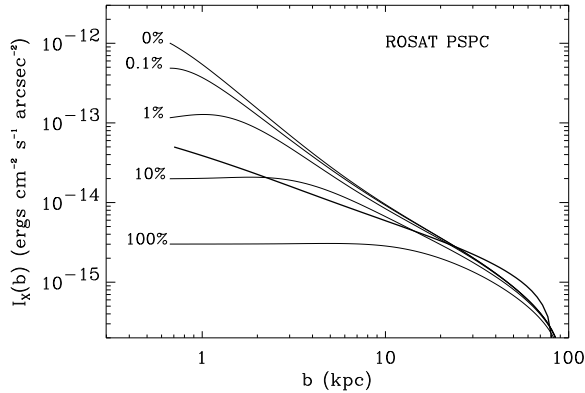}
\caption{The effect of internal X-ray absorption on the X-ray surface
brightness profile of a cooling flow (Wise \& Sarazin 1996).
The thin curves show the ROSAT PSPC profile for a
$\dot{M} = 300 \, M_\odot$ yr$^{-1}$, nearly homogeneous
$q=0.1$ model with various amounts of absorption (corresponding to the
labeled  fraction of the cooling gas going into the absorber).
The model labeled ``10\%'' has a spectrally determined excess column of
$\Delta N_H \approx 10^{21}$ cm$^{-2}$.
The heavy solid line represents the expected surface brightness
profile for an unabsorbed model with $\dot M(<r) \propto r$,
which is typical of the observed profiles.}
\label{fig:wise_sb}
\end{figure}

The most interesting result associated with the X-ray absorber
is its effect on the X-ray surface brightness profiles of cooling flows.
As shown in Figure~\ref{fig:wise_sb}, internal absorption flattens the
surface brightness profile of a cooling flow.
This occurs because both the absorber and emitter are concentrated to the
center of the cooling flow, and the absorber is thus particularly effective
at reducing the surface brightness in the center.
If the effects of the intrinsic absorption are ignored, this flattening
would be interpreted as evidence that the cooling flow gas is very
inhomogeneous.
For example, the cooling flow model assumed in Figure~\ref{fig:wise_sb} was a
nearly homogeneous model ($q = 0.1$ in the notation of
Wise \& Sarazin [1996]),
while the thick curve shows the surface brightness for very inhomogeneous
model with $\dot M(<r) \propto r$.

\subsection{What is the Excess Absorber?}
\label{sec:absorp_clouds}

In general, the cold material producing this absorption has not been
detected at non--X-ray wavelengths, despite considerable efforts
(e.g., McNamara \& Jaffe 1993;
Antonucci \& Barvainis 1994;
O'Dea et al.\ 1994).
The observational limits based on observations of H~I or CO
have become quite restrictive
(e.g., Voit \& Donahue 1995).
It has been suggested that the absorber might be
very cold molecular clouds
(Ferland et al.\ 1994),
or cold clouds in which all of the volatiles have frozen onto
dust grains
(Daines et al.\ 1994;
Fabian 1994).
However, the X-ray absorbing clouds should be detectable;
they absorb $\sim 3 \times 10^{43}$ ergs s$^{-1}$ of X-rays,
and must be re-radiating this luminosity at some wavelength.

A number of theoretical models have been constructed in order to
determine the physical properties of cold clouds in cooling flows
and decide whether the X-ray absorbing material should have been
detected in the existing H~I and CO observations
(Daines et al.\ 1994;
Ferland et al.\ 1994;
O'Dea et al.\ 1994;
Voit \& Donahue 1995).
These calculations have reached very different conclusions about
the viability of cold clouds in cooling flows.
It would be extremely helpful to understand the origin of this
discrepancy, and to reach a theoretical consensus as to the
physical state of X-ray absorbing clouds or other cold clouds
in cooling flows.
At the moment, we have no generally accepted theoretical
model for the possible origin of the X-ray absorber.

Current observations with ISO should provide important new information
about the nature of X-ray absorbing clouds or other cold clouds in
cooling flows.
ISO observations of the atomic fine structure lines
(e.g., [C~I], [O~I], and [Si~II]) will either detect or strongly
limit atomic gas as a source of the X-ray absorption, as these
lines are the primary coolants of atomic gas under most circumstances.
ISO observations of the continuum emission from cooling flows should
detect or limit the amount of X-ray absorption in dusty clouds.

\section{Magnetic Fields in Cooling Flows} \label{sec:magnet}

In addition to the thermal plasma, the intracluster medium contains
magnetic fields.
In clusters with diffuse radio emission
(e.g., Jaffe 1992),
X-ray limits on the amount of inverse Compton emission
give lower limits to the strength of the magnetic field
which are typically
$B \ga 0.1 \, \mu$G
(e.g., Rephaeli et al.\ 1987).
Faraday rotation measurements
towards background and cluster radio sources have also been used to
determine the intracluster magnetic field
(e.g., Kim et al.\ 1990).
The measured values of and upper limits on the
Faraday rotation are $RM \la 100$ rad m$^{-2}$, where $RM$ is the
rotation measure.
These observations are consistent with an
intracluster field strength and coherence length of roughly
$B \sim 1 \, \mu$G and $l_B \la 10$ kpc.
With this value for the field strength, the ratio of
magnetic to gas pressure is roughly $( P_B / P ) \la 10^{-3}$,
implying that the field is too weak to affect the dynamics of the
outer parts of clusters.

Although cluster magnetic fields may be generally weak, they are
enormously amplified by the compression and inflow in cooling flows
(Soker \& Sarazin 1990).
For frozen-in fields, the pressure associated with the magnetic field
increases dramatically;
e.g., $P_B \propto r^{-4}$ for homogeneous radial inflow.
Soker \& Sarazin showed that the fields should reach equipartition with the
thermal gas pressure within a typical radius of $\sim$20 kpc from the
center of the flow.
In the inner regions of cooling flows, the magnetic field
should then be very important dynamically.
The rapid amplification of the magnetic field in cooling flows also implies
a large increase in the rotation measure.  
Soker \& Sarazin (1990) showed that the resulting rotation measures
will be
\begin{eqnarray}
RM & \approx & 4000
\left( \frac{n_c}{3 \times 10^{-3} \, {\rm cm}^{-3}} \right)^{2/3}
\left( \frac{l_{Bc}}{10 \, {\rm kpc}} \right)^{1/2}
\nonumber  \\
&& \qquad
\times \left( \frac{{\dot M}_c}{100 \, M_\odot \, {\rm yr}^{-1}} \right)^{1/2}
\left( \frac{ T_c}{7 \times 10^7 \, {\rm K}} \right)^{1/2}
\, {\rm rad} \, {\rm m}^{-2} \, ,
\label{eq:rotm}
\end{eqnarray}
where $n_c$, $T_c$, ${\dot M}_c$, and $l_{Bc}$ are the electron density,
temperature, total cooling rate, and magnetic coherence length,
respectively, at the cooling radius. 
In the inner regions of the cooling flow, the magnetic coherence length
is still expected to be about 10 kpc.
Some initial MHD simulations of cluster cooling flow by
Christodoulou \& Sarazin (1996) have confirmed these results.

Often, the central galaxy in a cluster cooling flow is a radio galaxy,
and these radio sources have been used to search for Faraday rotation.
In all cases observed so far, the central radio sources in cluster cooling
flows have either very large Faraday rotations or depolarization.
Examples include M87/Virgo
(Owen et al.\ 1990),
Cygnus A
(Dreher et al.\ 1987),
Hydra A
(Taylor \& Perley 1993),
3C295
(Perley \& Taylor 1991),
A1795, A2199, A2052
(Ge 1991;
Ge \& Owen 1993),
A2029, and A4059
(Taylor et al.\ 1994).
These radio sources have rotation measures of
$RM \approx 10^3 - 2 \times 10^4$ rad m$^{-2}$,
which imply magnetic fields with
$B \ga 10 \, \mu$G and $l_B \sim 10$ kpc.
From a survey of Faraday rotations, Ge (1991) concluded that ``all
sources in the centers of strong cooling flows have high $RM$ ($\ga
1000$ rad m$^{-2}$),'' and that all other sources (in the centers of
clusters without cooling flows and in the outer regions of clusters
with cooling flows) have much smaller $RM$'s ($\la 100$ rad m$^{-2}$).
Taylor et al.\ (1994) found that the rotation measures were positively
correlated with the total cooling rates.
A number of cases have also been found of amorphous radio sources at the
centers of cooling flows which are highly depolarized;
PKS0745-191 (Baum \& O'Dea 1991), and
2A0335+096 (Sarazin et al.\ 1995a)
are the best studied and clearest cases.

These observations confirm the prediction that the magnetic fields
in cooling flows are strong.
Since the amplification of the field is due to compression and inflow,
the large rotation measures provide indirect evidence that gas is
indeed flowing into cooling flows.
The implied fields in the inner regions give magnetic pressures which
are comparable to the very high gas pressures;
thus, it is likely that magnetic fields affect the dynamics of the
gas in these regions.

One important question is the ultimate fate of the magnetic flux which
is advected into cooling flows.
If the flux were not removed by some process, the field would grow
to very large levels.
The magnetic fields might be convected out of the cluster center, or
might be destroyed by field line reconnection.
Physical arguments and initial numerical simulations suggest that
reconnection is more important in cooling flows
(Soker \& Sarazin 1990;
Christodoulou \& Sarazin 1996).

\section{Radio Sources in Cooling Flows} \label{sec:radio}

Most of the central galaxies in cluster cooling flows host
radio galaxies
(Burns 1996).
In fact, many of the most famous, nearby radio galaxies
(e.g., Virgo A, Perseus A, Cygnus A) are located in the centers of
cluster cooling flows.
(However, there do exist cases of moderately strong cooling
flows without a radio source at the center
[e.g., A376, A2319, A2141]).
Most of the radio sources associated with the central galaxies
in cluster cooling flows are FR~I (edge-darkened) sources;
an exception is the Cyg~A, which is an FR~II (edge-brightened)
source.
Recent observations of the FR~I radio sources in cluster cooling flows
suggest that they may be subdivided further into two separate morphologies.
Most of these radio sources show a radio jet or pair of jets which lead
from the nucleus to a pair of radio lobes.
I will refer to such sources as ``lobe-dominated sources.''
Examples of lobe-dominated sources in large cluster cooling
flows include Perseus~A, A1795, A2029, A2597, and A4059.

The second class of FR~I radio sources in cooling flows
are ``amorphous sources''
(Burns 1990;
Baum \& O'Dea 1991).
These seem to be less common that the lobe-dominated sources.
PKS0745-191 and 2A0335+096 are probably the best examples
(Baum \& O'Dea 1991;
Sarazin et al.\ 1995a).
In both of these radio sources, there is radio emission from
the galactic nucleus, but any jets are either very weak or
completely absent, even when the sources have been mapped with
a wide range of angular resolutions and at a wide range of radio
frequencies.
Most of the radio luminosity in these sources
comes from a extended of region diffuse, steep spectral index
emission.
There is no clear evidence for radio lobes or strongly directed
outflow of radio plasma.

A key clue to the dynamics of these radio sources comes from
comparing the pressure of the nonthermal radio emitting plasma
$P_{rad}$ with the pressure of the ambient, thermal, X-ray emitting gas, $P_X$.
We derive the radio pressure from synchrotron theory,
making the usual ``minimum energy''  assumptions.
The average X-ray pressure at the radius of the extended radio
emission (either the lobes or the amorphous emission) is
derived from the azimuthally averaged X-ray surface brightness.
We find that the X-ray and radio pressures are generally
in fairly good agreement (factor of three) for the
lobe-dominated sources.
For example, in A2597
the average radio pressure in the two lobes is
$P_{rad} \approx 1.1 \times 10^{-9}$ dyn cm$^{-2}$,
while the X-ray pressure (at a slightly larger effective radius
because of the small size of the radio source) is
$P_{X} \approx 0.5 \times 10^{-9}$ dyn cm$^{-2}$
(Sarazin et al.\ 1995b).
This suggests that the radio lobes are distinct from and confined
by surrounding thermal gas.
In the amorphous sources, the radio pressures are much smaller
than the X-ray pressures.
In 2A0335+096,
the average radio pressure in the diffuse emission region is
$P_{rad} \approx 2.7 \times 10^{-12}$ dyn cm$^{-2}$,
while the X-ray pressure in the same region (in projection) is
$P_{X} \approx 1.1 \times 10^{-10}$ dyn cm$^{-2}$.
A similar result is found in PKS0745-191
(Baum \& O'Dea 1991).
While it is possible that the disagreement between the radio
pressure and thermal pressure in these objects is due to a
failure of the minimum energy assumptions,
I believe that this discrepancy indicates that the radio
plasma is mixed with the thermal plasma (either on a fine or
coarse scale).
If the relativistic particles which produce the radio emission
occupy the same region as the thermal plasma, the partial pressure of
the radio plasma need not balance the pressure of
the thermal plasma.

\begin{figure}
\vspace{2.25in}
\includegraphics{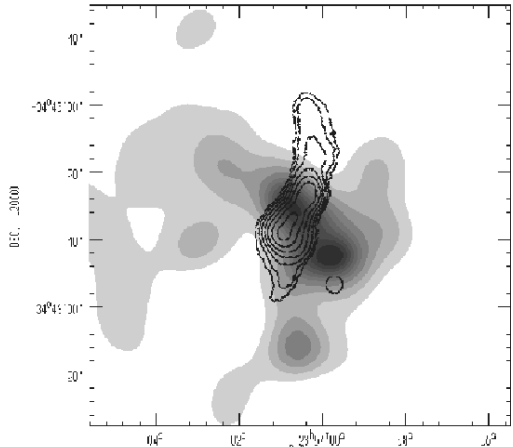}
\caption{Contours of the radio emission
from central galaxy in the cooling flow cluster A4059
are shown superposed on a greyscale representation of
the ROSAT HRI X-ray image
(Huang \& Sarazin 1996).}
\label{fig:a4059_x_radio}
\end{figure}

Another interesting result emerges if one compares the detailed
images of the central regions of cooling flows in radio and
in X-rays.
For the lobe-dominated sources, the X-ray emission and
the radio lobes appear to be anti-correlated.
That is, the projected region of the radio lobes is a region
of fainter X-ray emission, compared to the average X-ray surface
brightness at that radio.
The clearest example of this was shown in the ROSAT HRI X-ray
image of the Perseus cluster by
B\"ohringer et al.\ (1993).
However, we have found a similar effect in the ROSAT HRI
images of A1795, A2029, A2597, and A4059
(Sarazin et al.\ 1992, 1995b; Huang \& Sarazin 1996).
Figure~\ref{fig:a4059_x_radio} shows contours of the 
radio emission from the central radio source in
the cooling flow cluster A4059 superposed on the ROSAT HRI
X-ray image
(Huang \& Sarazin 1996).
We see that the X-ray emission is elongated ENE to WSW,
and that the radio lobes appear to be occupy regions of
lower X-ray brightness.

In the amorphous sources, the X-ray and radio emission occupy
the same projected regions.
If anything, the radio and X-ray emission appear to be
positively correlated
(e.g., Sarazin et al.\ 1995a).
This suggests that the radio emission and X-ray emission
come from the same volume of space.
Any process which compressed the thermal gas would increase
both the X-ray and radio emission, and this could produce
some level of correlation.

The polarization properties of these sources also connect the X-ray
emitting thermal plasma with the radio plasma.
As noted above (\S~\ref{sec:magnet}), the radio sources associated with
cluster cooling flows all show either large Faraday rotations or
depolarization.
In every lobe-dominated cooling flow source observed,
a very strong Faraday rotation
($RM \approx 10^{3} - 3 \times 10^{4}$ rad m$^{-2}$)
is observed.
On the other hand, both of the two amorphous sources
which have been observed (PKS0745-191, 2A0335+096)
showed complete depolarization
(Baum \& O'Dea 1991;
Sarazin et al.\ 1995a).

Large Faraday rotations (which imply the the polarization
vector undergoes many rotations) can only be produced if the
magnetized thermal plasma lies in front of the radio plasma.
If the thermal plasma and radio plasma are mixed (either
on a fine or coarse scale) the Faraday rotation will vary
along any line of sight through the radio source.
The differential Faraday rotation along each line of sight will
result in emission which is the superposition of all polarization
angles --- that is, unpolarized radiation.
Thus, the strong Faraday rotation seen in lobe-dominated cooling
flow radio sources indicates that the radio plasma has displaced
the X-ray emitting thermal plasma.
Conversely, the depolarization of the amorphous sources indicates
that the radio plasma and thermal plasma are mixed.

In summary, the comparison of X-ray and radio pressure,
of X-ray and radio images, and of polarization properties
are all consistent with a picture in which the lobes in
the lobe-dominated radio sources have displaced the surrounding
thermal gas, while the radio plasma in the amorphous sources
appears to be mixed with the X-ray emitting gas.

\acknowledgments

This work was supported by NASA Astrophysical\linebreak
Theory Program grant
NAG 5-3057, NASA ASCA grant NAG 5-2526, and NASA ASCA grant 5-3308.
I would like to thank Noam Soker and others at Oranim for the wonderful
job they did in organizing this very useful meeting.

\end{document}